\documentstyle[aps,prl,manuscript]{revtex}
\begin{document}

\draft


\title {Screening effects in the electron-optical phonon interaction}
\author {Michael Reizer}
\address{5614 Naiche Rd. Columbus OH 43213}
\date{\today}
\maketitle
\begin{abstract}
\noindent

We show that recently reported \cite{AS} unusual 
hardening of optical phonons
renormalized by the electron-phonon interaction \cite{AS} is due to 
the neglect of screening effects. When the electron-ion interaction
is properly screened optical phonons soften in three dimension.
It is important that for short-wavelength optical phonons  
screening is static while for long-wavelength optical phonons 
screening is dynamic. In two-dimensional
and one-dimensional cases due to crossing of the nonperturbed 
optical mode with gapless plasmons the spectrum of 
renormalized optical phonon-plasmon mode shows split 
momentum dependence.
\end{abstract}

\bigskip


The the electron-phonon interaction in metals was historically
one of the first many-body problem studied with the field theory methods
\cite{M}. The theory initially developed for the acoustic phonons
was very successful due to the Migdal theorem \cite{M},
which states that the vertex corrections are small.
Naive extension of the theory for optical phonons
immediately meets some difficulties. 

Let's consider e.g. the attenuation of optical phonons,
$\gamma_q=-2\lambda\omega_0{\rm Im}\Pi^R(q,\omega)$, where
$\lambda$ is the dimensionless coupling constant and $\Pi^R$ is
the retarded polarization operator. Without vertex corrections,
the polarization operator is
\begin{eqnarray}
\label{1}
\Pi_0(q,\omega)={\nu\over 3}\biggl({qv_F\over \omega}\biggr)^2,
\ \ \ qv_F<<\omega;\nonumber \\ 
\Pi_0^R(q,\omega)=-\nu\biggl(1+i{\pi\omega\over 2qv_F}\biggr),
\ \  \omega<<qv_F, 
\end{eqnarray}
where $\nu$ is the two-spin electron density of states and $v_F$ is
the Fermi velocity.
We see that for long-wavelength optical phonons 
$\omega>qv_F$ the attenuation is absent and near the threshold
$\omega\lesssim qv_F$ the attenuation is anomalously large so
the optical phonon is a poor defined elementary excitation. This result
is certainly unphysical and contradict experimental data,
the optical phonon attenuation is measured by Raman light scattering.
The contradiction was resolved in 
\cite{R1} and \cite{R2}. It was shown that it is necessary to
take into account that the screening of the electron-ion potential 
is different for long-wavelength and short-wavelength optical phonons.
Besides that the attenuation of long-wavelength optical phonons 
requires the two-phonon
processes to be included, which are beyond the Migdal theorem. 
The obtained results are in reasonable agreement with the experiment.

The other difficulty arises for renormalization of the
optical phonon. 
According to Ref. \cite{M} the first correction to the electron-phonon
vertex function $\Gamma^{(1)}$ is proportional to
\begin{eqnarray}
\label{2}
\Gamma^{(1)}(q,\omega)\sim-i\int d\Omega
{\omega\over \omega-{\bf q}\cdot{\bf v}}
\Gamma(q,\omega),                                               
\end{eqnarray}
where $\Gamma(q,\omega)$ is the bare electron-phonon vertex function,
and $\Omega$ is the electron angular variables.
Really for the acoustic phonons $\omega=\omega_q=uq$, 
the correction is small, $\Gamma^{(1)}\sim (u/v_F)\Gamma<<\Gamma$, 
where $u$ and $v_F$ are the sound velocity.

However the Migdal theorem does not hold for long-wavelength optical
phonons. Indeed  Eq. (1) shows that for long-wavelength optical phonons,
$\omega_q\approx\omega_0$, $qv_F<<\omega_0$,
the vertex correction is not small, $\Gamma^{(1)}\sim\Gamma$,
and should be included in any calculations.
For the problem of renormalization of optical phonons 
higher order terms in the electron-phonon interaction were studied in
Ref. \cite{ES} and recently reexamined in \cite{AS}. 
Calculating the second order correction to 
the phonon polarization operator, the authors of Ref. \cite{AS} found the
following renormalized dispersion of the long-wavelength optical phonon
\begin{eqnarray}
\label{3}
\omega_q\approx\omega_0+{\lambda\over 3}\biggl({qv_F\over \omega_0}
\biggr)^2(1-2\lambda\ln2),                                                     \end{eqnarray}
where $\lambda$ is a dimensionless coupling constant.
As was correctly mentioned in \cite {AS} hardening of the phonon spectrum
shown by Eq. (2) contradicts theoretical expectation of the Migdal
theory and it also contradicts the experimental results.
 
The purpose of the present paper is to show that unusual renormalization  
of the optical phonon is due
to the neglect of screening effects. We will show that with 
the screening effects taken 
into account the spectrum of the optical phonons is softened.

The screened electron-phonon vertex function for longitudinal phonons is
\begin{eqnarray}
\label{4}
\Gamma_s(q,\omega)={\Gamma_0(q)\over \epsilon(q,\omega)}, \nonumber \\ 
\Gamma_0(q)=-i{\bf q}\cdot{\bf e}_\lambda
{4\pi e^2Z\over q^2}
\biggl({N\over 2M\omega_q^0}\biggr)^{1/2},                          
\end{eqnarray}
where $Z$ is the ion valence,
$M$ is the ion mass, $N$ is the number of unit cells, 
${\bf e}_\lambda$ is the phonon polarization vector.

The dielectric function, $\epsilon(q,\omega)$, 
in the random phase approximation is expressed trough 
the Coulomb potential
$V_q=4\pi e^2/q^2$ and the electronic polarization operator 
by the equation $\epsilon(q,\omega)=1-V_q\Pi(q,\omega)$.
According to \cite{AS} the first order vertex
and self-energy corrections from the electron-phonon
interaction to the polarization operator in the long-wavelength limit,
$qv_F<<\omega$ is
$\Pi_1(q,\omega)=-2\lambda\ln 2\Pi_0(q,\omega)$.
The correction is due to virtual optical or acoustic phonons with
large momentum transfer. Thus the full polarization operator for
long-wavelength optical phonons is $\Pi(q,\omega)=\Pi_0(q,\omega)+
\Pi_1(q,\omega)=(1-2\lambda\ln 2)\Pi_0(q,\omega)$.

The Dyson equation for the renormalized longitudinal optical phonon 
frequency $\omega_q$ is
\begin{eqnarray}
\label{5}
\omega_q^2=(\omega_q^0)^2+2\omega_q^0|\Gamma_0(q,\omega_q^0)|^2
{\Pi(q,\omega_q)\over\epsilon(q,\omega_q)},
\end{eqnarray}
where $\omega_q^0$ is the nonperturbed optical phonon frequency.
We screened only one vertex, (see e.g. \cite{S}).

For short-wavelength optical phonons the Migdal theorem holds
and we neglect the term $\Pi_1$, then
\begin{eqnarray}
\label{6}
\omega_q^2=
(\omega_q^0)^2+{q^2Z^2N\over M}{V_q^2\Pi_0(q,\omega_q)\over 
1-V_q\Pi_0(q,\omega_q)}.                                            
\end{eqnarray}
where $\omega_q^0$ is the nonperturbed optical phonon frequency and
according to \cite{S} we screened only one vertex.
For acoustic phonons Eq. 6 coincides with the Bohm-Staver equation for
the jellium model (see e.g. \cite{S}) where $\omega_q^0$ corresponds
to ionic plasma frequency $\Omega_{pl}^2=(Zm/M)\omega_{pl}^2$, where
$\omega^2_{pl}=V_q(qv_F)^2\nu/3=4\pi e^2n/m$ is the electronic plasma 
frequency, and $n=ZN$ is the electron density. Assuming that
condition $\omega_q<<v_Fq$ is satisfied, 
Eq. 6 leads to the renormalized acoustic phonon frequency 
$\omega_q=uq$, where $u^2=(Zm/3M)v_F^2$. 
Note that there is no optical phonons in the jellium model,
thus we assume that nonperturbed optical phonon frequency
$\omega_q^0$ is calculated within some other physically reasonable model,
e.g. within the uniform background model \cite{C}.
For short-wavelength optical phonons we have
\begin{eqnarray}
\label{7}
\omega_q^2=(\omega_q^0)^2-{Zm\over M}(\omega_{pl})^2{\kappa^2\over 
q^2+\kappa^2},  \ \ \           \omega_q<<qv_F,
\end{eqnarray}
For  long-wavelength optical phonons, $qv_F<<\omega_q$, 
we keep $\Pi_1$ term, thus
\begin{eqnarray}
\label{8}
\omega_q^2=(\omega_q^0)^2+{q^2Z^2N\over M}{V_q^2\Pi(q,\omega_q)\over 
1-V_q\Pi(q,\omega_q)}.
\end{eqnarray}
Eq. (8) is transformed to
\begin{eqnarray}
\label{9}
\omega_q^2=(\omega_q^0)^2+{Zm\over M}{(\omega_{pl})^4A
\over \omega_q^2-\omega_{pl}^2A},\ \ A=1-2\lambda\ln 2.
\end{eqnarray}
In good metals $\omega_q<<\omega_{pl}$, and if we take into account
dispersion of the plasma frequency $\omega_{pl}^2(q)=4\pi e^2n/m+3/5(qv_F)^2$,
then Eq. (9) shows softening of the nonperturbed optical 
phonon frequency $\omega_q^0$. 
Note also that for short-wavelength optical
phonons the screening of the electron-ion interaction is static. For 
long-wavelength optical phonon the screening is dynamic,
the corresponding dielectric function
$\epsilon(q,\omega_q)=1-(\omega_{pl}/\omega_q)^2$ for $\omega_q<\omega_{pl}$
is negative (the antiscreening effect), which corresponds to 
out-of-phase oscillations of electrons and ions.

The obtained results may be easily applied to a semiconductor with
the polarization
coupling of electrons with optical phonons. Corresponding
electron-phonon vertex function is \cite{Rid}
\begin{eqnarray}
\label{10}
\Gamma_0(q)=-i{\bf q}\cdot{\bf e}_\lambda BZ V_q
\biggl({N\over 2M\omega_q^0}\biggr)^{1/2}.                          
\end{eqnarray}
We see that the main difference is the appearance of the coupling
constant $B$ which is defined in \cite{Rid}. Thus Eq. (9) is valid
with simple substitution $Z\to B^2Z$. Therefore for $\omega_q<<\omega_{pl}$
polarization electron-optical phonon interaction leads to softening
of optical phonons. In the opposite case, $\omega_q>>\omega_{pl}$ 
the renormalization of the optical phonon spectrum is negligible,
\begin{eqnarray}
\label{11}
\omega_q^2=(\omega_q^0)^2-{ZB^2Am\over M}(\omega_{pl})^2{\omega_{pl}^2
\over \omega_q^2}\approx (\omega_q^0)^2.
\end{eqnarray}

Situation becomes more interesting in the low-dimensional electron system
because in two- and one-dimensions electronic collective excitations, 
the plasmons, are gapless and
crossing of the nonperturbed optical mode with the plasmon may take place,
which in turn leads to qualitatively new dispersion of the
optical phonon in systems of restricted dimensionality.

In two and one dimensions
\begin{eqnarray}
\label{12}
\omega_{pl2}^2(q)= {1\over 2}\kappa_2v_F^2q, \ \  
\kappa_2=2\pi e^2\nu_2,\ \ \  \nu_2=m/\pi,                    
\end{eqnarray}
\begin{equation}
\label{13} 
\omega_{pl1}^2(q)={2v_Fq^2\over \pi}2e^2[\ln(1/qa)+1.972...], qa<<1,  
\end{equation}                                                              
where $a$ is a width of a wire.
The solution of Eq. (9) for deformation coupling and without renormalization 
constant $A$ is
\begin{eqnarray}
\label{14}
\omega_q^2={1\over 2}[\omega_0^2+\omega_{pli}^2(q)]\nonumber \\
\pm{1\over 2}\biggl[[\omega_0^2-\omega_{pli}^2(q)]^2+
{4Zm\over M}\omega_{pli}^4(q)\biggr]^{1/2},                    
\end{eqnarray}
where $\omega_{pli}$ equals to $\omega_{pl2}$ and $\omega_{pl1}$ in 
two and one dimensions correspondingly.
Near the spectrum crossing $\omega_{pli}(q)=\omega_0$ the solution is
\begin{eqnarray}
\label{15}
\omega_q^2={1\over 2}\omega_0^2+\omega_{pli}^2(q)
\biggl[{1\over 2}\pm{Zm\over M}\biggr].                        
\end{eqnarray}
Equation (15) shows typical spectrum splitting between
the optical phonon and the plasmon.
Unfortunately in such low-dimensional electron systems as GaAr
heterostructures the optical frequency is much higher than plasmon modes.
The situation described above may take place in some other
low-dimensional systems such as conducting polymers or
the systems close to structural instability when the optical mode
becomes soft for selective wave vectors.

In conclusion we have shown that physically expected softening
of optical phonons due to the electron-phonon interaction is obtained
when the electron-ion interaction is screened properly.
The screening effect is different for long-wavelength and 
short-wavelength optical phonons. In low-dimensional
electron systems the optical mode may cross the
gapless plasmon mode, as a result new optical phonon-plasmon
collective mode has a split momentum dispersion.


\begin{references}

\bibitem{M}. A. B. Migdal, Zh. Eksp. Teor. Fiz. {\bf 34}, 1438 (1958).
\bibitem{R1}. M. Yu. Reizer, Solid. State. Comm. {\bf 44}, 237 (1982).
\bibitem{R2}. M. Yu. Reizer, Phys Rev. B {\bf 38}, 10398 (1988). 
\bibitem{ES}. S. Engelsberg and J. R. Schrieffer, Phys. Rev. {\bf 131},
993 (1963).
\bibitem{AS}. A. S. Alexandrov and J. R. Schrieffer, Phys. Rev. B, {\bf 56},
13731 (1997).
\bibitem{S}. J. R. Schrieffer, {\it Theory of Superconductivity}
(Addison Wesley, Redwood City, 1988). Chapter 6.
\bibitem{C}. C. B. Clark, Phys. Rev. {\bf 109}, 1133 (1958). 
\bibitem{Rid}. B. K. Ridley, {\it Quantum processes in semiconductors},
(Clarendon Press Oxford, 1988). Chapter 3. 
\end{references}
\end{document}